\documentclass{amsproc}

\theoremstyle{definition}

\theoremstyle{remark}

\numberwithin{equation}{section}

\begin{document}
\begin{flushright}
TIFR/TH/01-24\\
hep-th/0107087\\
\end{flushright}
\vspace{1truecm}

\title{Ramond-Ramond Couplings of Noncommutative Branes}

\author{Sunil Mukhi}
\thanks{Based 
on an invited talk given by Nemani V. Suryanarayana at Strings 2001,
Mumbai, January 2001\cite{talklink}.}

\author{Nemani V Suryanarayana} 
\address{Tata Institute of Fundamental
Research, Homi Bhabha Rd, Mumbai 400 005, India}
\email{mukhi@tifr.res.in, nemani@tifr.res.in}


\begin{abstract}
We obtain the couplings of noncommutative branes of type II string
theories to constant Ramond-Ramond backgrounds, for BPS as well as
non-BPS branes, in the background-independent description. For the BPS
branes, we also generalize these couplings to other descriptions, and
thereby argue their equivalence to the known couplings in the
commutative description. The first part is a review of earlier work
while the second part contains some additional observations.
\end{abstract}

\maketitle

\section{Introduction}
Much insight has been gained into the dynamics of branes in string
theory using noncommutativity \cite{noncommrefs, SW}. In the presence
of a constant 2-form B-field, one finds that the world-volume action
of a D-brane can be described either in terms of commutative or
noncommutative variables. Using the continuous description parameter
$\Phi$, one can actually interpolate between the two types of
descriptions. In Ref.\cite{SW} the DBI action for a D-brane in the
presence of a constant B-field was proposed in a general description
and was shown to be equivalent to the commutative one.

In recent times noncommutativity has also proved useful in
understanding the issue of tachyon condensation in the case of
unstable branes \cite{gmsone}-\cite{sennew}. On an unstable non-BPS
D-brane one gets a noncommutative field theory involving tachyonic
scalars, which generically admits static solitonic solutions over
which tachyon condensation can occur representing brane decay.

A distinguished feature of D-branes in Superstring theories is that
they couple to the Ramond-Ramond fields given by the Chern-Simons
terms. With the recent interest in the noncommutative descriptions of
D-branes, a natural question to ask is: How are these couplings
described in the noncommutative language? Here we try to address this
question for the case of constant Ramond-Ramond fields for both BPS as
well as non-BPS branes of type II string theories.

We first review the noncommutative DBI action for a single (Euclidean)
Dp-brane. Thinking of a Dp-brane as a classical configuration of
infinitely many D(-1)-branes, we obtain the noncommutative DBI action
in $\Phi = -B$ description \cite{cornalba, seibnew}. We then follow
the same prescription to obtain the Noncommutative Chern-Simons terms
in the background-independent $\Phi = -B$ description. Next we propose
analogous couplings on non-BPS branes. One key property of
noncommutative solitons is that the tachyon condensation over them can
produce $N$ coincident lower dimensional D-branes starting from a
single higher dimensional noncommutative brane. In recent times it has
emerged that collections of $N$ D-branes have extra commutator
couplings in their world-volume theory, to Ramond-Ramond potentials,
that do not exist for a single D-brane \cite{myers, twosens}. We show
that such terms can be obtained by producing the lower dimensional
branes via tachyon condensation over an appropriate noncommutative
soliton starting from our proposed answer for the Ramond-Ramond
couplings. Finally we propose generalisations of the noncommutative
Chern-Simons terms on the BPS branes to other descriptions and show
that these couplings are equivalent to the commutative ones in the DBI
approximation (upto total derivatives). We work with Euclidean branes
with an even number of world-volume directions. In what follows we set
$2\pi\alpha' = 1$. The talk on which this article is
based\cite{talklink} contained a
review of Ref.\cite{nccs}, along with some new observations about
description dependence. Below, in the conclusions, we also briefly
review our subsequent work, Refs.\cite{smnvstwo, sdsmnvs}.


In the presence of constant NS-NS B-field the dynamics of a
(Euclidean) Dp-brane can be described by the following DBI
action\cite{SW}:
\begin{eqnarray}
\hat{S}_{DBI} &=& \hat{T}_p \int d^{p+1} x~
\sqrt{\hbox{det}\big(G_{ij} + \hat{F}_{ij} + \Phi_{ij}\big)} \cr 
\end{eqnarray}
where $\hat T_p = (2\pi)^{\frac{1-p}{2}}/{G_s}$ and the field strength
is 
\begin{equation}
\hat F_{ij} =
\partial_i {\hat A}_j - \partial_j {\hat A}_i - i [\hat A_i, \hat
A_j]_{*}
\end{equation}
The products of fields appearing in this equation are understood to be
$*$ products given by:
\begin{equation}
f(x) * g(x) \equiv e^{\frac{i}{2}\theta^{ij}\partial_i \partial'_j} f(x)
g(x')\big|_{x=x'} 
\end{equation}
The parameters $G_{ij}, \Phi_{ij}, G_s$ and $\theta^{ij}$ are given in 
terms of the commutative variables $g, B, g_s$ by:
\begin{equation}
\label{defs}
\frac{1}{G + \Phi} = -\,\theta + \frac{1}{g+B}, \qquad\qquad G_s = g_s
\left(\frac{\hbox{det}(G+\Phi)}{\hbox{det}(g+B)}\right)^{\frac{1}{2}}
\end{equation}
where $\Phi$ denotes the freedom in description. Three choices of
$\Phi$ are of particular interest. They are $\Phi_{ij}=B_{ij}$,
$\Phi_{ij} = 0$ and $\Phi_{ij} = - B_{ij}$. The action in terms of
commutative variables with ordinary products corresponds to
$\Phi_{ij}=B_{ij}$ description. $\Phi_{ij} = - B_{ij}$ is the
background independent description. In this description, we have
\begin{equation}
\theta^{ij} = (B^{-1})^{ij},\qquad\qquad
G_{ij} = - B_{ik}g^{kl}B_{lj},\qquad\qquad
G_s = g_s \sqrt{\frac{\hbox{det}\,B}{\hbox{det}\,g}}
\end{equation}
Working in $\Phi = -B$ description  $\hat S_{DBI}$ can be put 
in the form:
\begin{eqnarray}
\hat S_{DBI}=T_p 
\int d^{p+1} x~ \frac{\hbox{Pf}\,Q}{\hbox{Pf}\, \theta} 
~\sqrt{\hbox{det}\big(g_{ij} + (Q^{-1})_{ij}\big)}
\end{eqnarray}
Where $Q^{ij} = \theta^{ij} - \theta^{ik}\,\hat F_{kl}\,\theta^{lj} =
-i[X^i, X^j]$ with $X^{i} = x^i + \theta^{ik}\hat A_k$ and $T_p =
(2\pi)^{\frac{1-p}{2}}/g_s$

We re-express this action in terms of a trace Tr over a Hilbert space.
Using
\begin{equation}
\int d^{p+1}x \rightarrow \hbox{Tr} (2\pi)^{\frac{p+1}{2}} \hbox{Pf}\,\theta
\end{equation}
we get
\begin{eqnarray}
\hat{S}_{DBI} = \frac{2\pi}{g_s} 
\hbox{Tr}\left[ \hbox{Pf}\, Q \sqrt{\hbox{det}\big(g_{ij} +
(Q^{-1})_{ij}\big)}\right]
\end{eqnarray}

This action can be thought of as the action for infinitely many D(-1)
- branes in a classical configuration. To see this we can start
with the nonabelian DBI action for N D(-1)-branes (with N $\rightarrow
~\infty$) \cite{myers}:
\begin{equation}
\hat{S}_{DBI} = \frac{2\pi}{g_s} 
\hbox{Tr}\left[ \sqrt{\hbox{det}\big(\delta_i^j - i g_{ik}[X^k,
X^j]\big)} \right]
\end{equation}
Now consider the solution corresponding to a Dp-brane:
\begin{eqnarray}
X^i = x^i \qquad\quad \hbox{for} \quad i &=& 1, 2,..., p+1 \cr
X^j = 0   \qquad\quad \hbox{for} \quad j &=& p+2,..., 10.\cr
\end{eqnarray}
such that
\begin{equation}
[x^i, x^j] = 
i\,\theta^{ij} \equiv (B^{-1})^{ij}
\end{equation}
Consider fluctuations around this classical solution:
\begin{eqnarray}
X^i &=& x^i + \theta^{ik}\hat A_k \cr
X^j &=& \phi^j 
\end{eqnarray}
Substituting these in the action for the D(-1)-branes gives
the noncommutative DBI action of a Dp-brane in $\Phi = -B$
description. We follow the same prescription to obtain the
Ramond-Ramond couplings on noncommutative BPS branes.

\section{Chern-Simons couplings on BPS noncommutative branes}

For this we start with the nonabelian Chern-Simons action for N D(-1)-branes:
(N $\rightarrow ~\infty$) \cite{myers} (for earlier work on
Ramond-Ramond couplings in the matrix model, see Ref.\cite{morariu})
\begin{equation}
S_{cs} = \frac{2\pi}{g_s} \hbox{Tr}\left[ e^{i(\hbox{i}_X
\hbox{i}_X)}\sum_n C^{(n)}\right]
\end{equation}
For example let us take the coupling to the RR 10-form  $C^{(10)}$:
\begin{eqnarray}
S_{cs} &=& \frac{2\pi}{g_s}\hbox{Tr}\left[ {\frac{i^5}{5!}}
X^{i_{10}}X^{i_9}...\, X^{i_1} C^{(10)}_{i_1i_2...i_{10}}\right] \cr
&=& -\frac{2\pi}{g_s}\hbox{Tr}\left[{\frac{1}{5! 2^5}}(i[X^{i_1},
X^{i_2}])...(i[X^{i_9},X^{i_{10}}]) C^{(10)}_{i_1i_2...i_{10}}\right]
\end{eqnarray}
To obtain the noncommutative Chern-Simons terms for a Dp-brane, we
substitute the solutions corresponding to a Dp-brane along with the
fluctuations into the action above. In the case of a D9-brane this
gives rise to:
\begin{eqnarray}
\hat S_{cs}^{(D9)} &=&
\frac{2\pi}{g_s}\hbox{Tr}\left[{\frac{1}{5!2^5}}
\epsilon_{i_1i_2...i_{10}} Q^{i_1i_2}...\,Q^{i_9i_{10}}
{\frac{1}{10!}}\epsilon^{j_1j_2...j_{10}}C^{(10)}_{j_1j_2...j_{10}}
\right]\cr
&=& \frac{2\pi}{g_s}\hbox{Tr}\left[\hbox{Pf}\,Q \,C^{(10)} \right]
\end{eqnarray}
Converting into an integral:
\begin{eqnarray}
\hat S_{cs}^{(D9)} = T_9\int{{d^{10}x}
{\frac{\hbox{Pf}\,Q}{\hbox{Pf}\,\theta}}\left[{\frac{1}{10!}}
\epsilon^{j_1j_2...j_{10}}C^{(10)}_{j_1j_2...j_{10}}\right]}
\end{eqnarray}
where again $Q = \theta - \theta \hat F \theta$.
For obtaining lower form couplings, we repeat the above exercise
starting with that form coupling to infinite D(-1)-branes.
Using the identity:
\begin{eqnarray}
&&\frac{\hbox{Pf}\,Q}{2^r r!(2n-2r)!}\,\epsilon^{i_1i_2\cdots i_{2n}}
Q^{-1}_{i_1i_2}\cdots Q^{-1}_{i_{2r-1}i_{2r}}
C^{(2n-2r)}_{i_{2r+1}\cdots i_{2n}} \cr
&& = 
\frac{(-1)^r}{2^{n-r}(n-r)!}\, Q^{i_{2r+1}i_{2r+2}}\cdots
Q^{i_{2n-1}i_{2n}} C^{(2n-2r)}_{i_{2r+1}\cdots i_{2n}}
\end{eqnarray}
the result can be put in the following form:
\begin{equation}
\hat{S}_{cs}^{(D9)} = T_9 \int_x \frac{\hbox{Pf}\, Q}{\hbox{Pf}\,\theta}
\sum_n C^{(n)}~ e^{Q^{-1}}
\end{equation}
For lower dimensional branes one can consider coupling to higher
RR-forms as well. In this case one has to again start with the
nonabelian Chern-Simons terms on N D(-1)-branes and substitute the
appropriate brane solution along with the fluctuations. For
illustration let us start with the case of a Euclidean D1-brane
coupling to the RR 4-form in type IIB. The coupling of $N$
D-instantons to the RR 4-form is
\begin{equation}
\hbox{Tr}\,\Big( {\frac{1}{ 2! 2^2}}\left(-i[\phi^{i_1},\phi^{i_2}]\right)
\left(-i[\phi^{i_3},\phi^{i_4}]\right) C^{(4)}_{i_1 i_2 i_3 i_4}
\Big)
\end{equation}
Here $\phi^i$ represent all 10 transverse scalars to the
D-instantons. Now insert $\phi^1=X^1,~\phi^2=X^2$. The remaining
$\phi^i$ are renamed as $\hat{\Phi}^a$, they represent the scalars
transverse to the noncommutative D1-brane. Thus we find the coupling:
\begin{equation}
\frac{1}{2! 2}\,\epsilon_{ij}\,\hbox{tr}\Big(
\big(-i[X^i,X^j]\big) \big(-i[\hat{\Phi}^a,\hat{\Phi}^b]\big)
- \big(-i[X^i,\hat{\Phi}^a]\big) \big(-i[X^j,\hat{\Phi}^b]\big)\Big)
C^{(4)}_{12ab} 
\end{equation}
Making the replacements
\begin{eqnarray}
-i[X^1,X^2] &=& Q^{12} = \theta^{12}(1+\theta^{12}\hat F_{12})\cr 
[X^i,\hat{\Phi}^a] &=& i\theta^{ij} D_j\hat{\Phi}^a\cr
\end{eqnarray}
the operator turns into:
\begin{equation}
\theta^{12}\Big( (1+\theta^{12}\hat
F_{12})\big(-i[\hat{\Phi}^a,\hat{\Phi}^b]
\big) + \theta^{ij}D_j\hat{\Phi}^a D_i\hat{\Phi}^b \Big) 
\end{equation}
Therefore we seem to have a non-vanishing coupling of a single
noncommutative D1-brane to a 4-form RR field. However it is easy to
see from the last expression that this coupling is zero in the DBI
approximation (neglecting ${\mathcal O}(\partial \hat F)$ and
${\mathcal O}(\partial^2 \hat \Phi)$ terms). Similarly one can find
the coupling of a noncommutative brane of any dimension to any RR-form. 

\section{Noncommutative Chern-Simons terms on non-BPS branes}

Type II theories have unstable Dp-branes, where p is odd for type IIA
and even for type IIB. Like the BPS branes, these branes also couple
to Ramond-Ramond forms \cite{sencs, billo, kennedy}. These terms on a
single unstable brane in commutative description are given by:
\begin{equation}
S_{cs} = {\frac{T_{p-2}}{2T_{min}}}\int_x 
dT~\sum_n C^{(n)} e^{F+B}
\end{equation}
Where $T_{min}$ is the value of the tachyon at the minimum of the
tachyon potential $V(T)$. We propose that the Chern-Simons action on
the Euclidean D9 brane of type IIA in the noncommutative description
is:
\begin{equation}
\hat{S}_{cs} = \frac{T_{8}}{2T_{min}}\int_x \frac{\hbox{Pf}\,
Q}{\hbox{Pf}\, \theta} ~{\mathcal{D}}T~ \sum_n C^{(n)}~e^{Q^{-1}}
\end{equation}
where ${\mathcal{D}}_i T = -i (Q^{-1})_{ij}[X^j, T]$. A non-trivial
check of this action is the following. Consider a noncommutative
soliton which represents the decay of a (Euclidean) D9 brane into N
coincident D(-1)-branes. Condensing the tachyon over that soliton we
will get N coincident non-BPS D(-1)-branes, which carry Myers type
couplings to the RR-forms \cite{twosens}. Substituting the solitonic
solution along with fluctuations around it should give us the
nonabelian Chern-Simons action on these branes. It is easy to verify
that the proposed action does pass this check. The action for other
unstable branes can be obtained by taking appropriate noncommutative
soliton representing lower branes and substituting the solution along
with fluctuations.

For illustration let us do the tachyon condensation over the following
soliton, which is supposed to represent the decay of a D9-brane into N
D7-branes.
\begin{eqnarray}
T_{cl} &=& T_{max} P_N + T_{min}(1-P_N)\cr
X^i_{cl} &=& P_N x^i\quad {\rm for}\quad i= 1,2,\ldots,8\cr
X^i_{cl} &=& 0 \quad {\rm for}\quad i= 9,10\cr
\end{eqnarray}
This solution has the property that $[T_{cl}, X^{i}_{cl}]
=0$. Substituting this solution along with the fluctuations into the
$C^{(9)}$ coupling in the proposed action for the unstable D9-brane,
we get
\begin{eqnarray}
\hat{S}^{UD7}_{CS} &=& \frac{T_{6}}{2T_{min}}{\hbox{tr}}_N 
\int_x (-i)~[\delta X^9, \delta X^{10}]~ \cr
&&(-i)(Q_{cl}^{-1})_{12}
~[X^2_{cl},\delta T]~ C^{(9)}_{2 3 \ldots 10}
\end{eqnarray}
And also
\begin{eqnarray}
\hat{S}^{UD7}_{CS} &=& \frac{T_{6}}{2T_{min}}{\hbox{tr}}_N\int_x 
(-i)[\delta X^{10},\delta T]~ C^{(9)}_{12\ldots 8,10}\cr
\end{eqnarray} 
where $\hbox{tr}_N$ denotes the trace over the $N\times N$ Chan-Paton
matrices. These are actually the two kinds of couplings that exist on
a stack unstable branes \cite{twosens}.

\section{General $\Phi$}

Here we propose the noncommutative Chern-Simons couplings on a BPS
brane in other descriptions and use them to argue the equivalence of
the actions in various descriptions in the DBI approximation, on the
lines of Ref.\cite{SW}. As we have seen the RR couplings to a
noncommutative Dp-brane are given by:
\begin{equation}
\hat{S}_{CS} = T_p \int_x {\sqrt{\hbox{det}(1-\theta \hat F)}}
\sum_nC^{(n)}e^{Q^{-1}}
\end{equation}
We propose that the answer in all other descriptions is given by (see
also \cite{micheliutwo}):
\begin{equation}
\hat{S}(\Phi)_{CS} = T_p \int_x {\sqrt{\hbox{det}(1-\theta \hat F)}}
\sum_nC^{(n)}e^{B+\hat F(1-\theta\hat F)^{-1}}
\end{equation}
where $\theta$ is the noncommutativity parameter for the corresponding
description, as given in Eq.(\ref{defs}). To verify this we consider
the variations of the action with respect to $\theta$ and show that
these variations vanish up to total derivatives and
${\mathcal{O}}(\partial{\hat F})$ terms. First we illustrate this for
the case of the top form coupling. The variation of $\hat F$ is
\cite{SW}
\begin{equation}
\delta {\hat F}_{ij}(\theta) = \delta\theta^{kl}\left[{\hat
F}_{ik}{\hat F}_{jl} -\frac{1}{2}{\hat A}_k(\partial_l {\hat F}_{ij} +
{\hat D}_l {\hat F}_{ij})\right] + {\mathcal{O}}(\partial{\hat F}) 
\end{equation}

Since we are going to ignore derivatives of $\hat F$, we drop the $*$
products between $\hat F$'s but leave them in the definitions of $\hat
F$ and in $\hat D_l \hat F_{ij}$. Keeping the closed string RR field
constant the variation of the action is given by:
\begin{eqnarray}
\delta \left[{\sqrt{\hbox{det}(1-\theta \hat F)}}\right]  &=&
-\frac{1}{2} {\sqrt{\hbox{det}(1-\theta \hat F)}}~ \hbox{Tr}\left[
{\frac{1}{1- \theta
\hat F}} \delta \theta \hat F + \frac{1}{1- \theta \hat F} \theta
\delta \hat F \right] \cr
&=& -\frac{1}{2} {\sqrt{\hbox{det}(1-\theta \hat F)}}~ \hbox{Tr} \Big[
\frac{1}{1- \theta
\hat F} \delta \theta \hat F + {\left({\frac{1}{1- \theta \hat
F}}\right)}^{j}~_{m} \cr
&&~~~~~\theta^{m i}\delta\theta^{kl}
\big({\hat
F}_{ik}{\hat F}_{jl} -\frac{1}{2}{\hat A}_k(\partial_l {\hat F}_{ij} +
{\hat D}_l {\hat F}_{ij})\big)\Big] + {\mathcal{O}}(\partial{\hat F}) \cr 
&=& -\frac{1}{2}{\sqrt{\hbox{det}(1-\theta \hat F)}}~\hbox{Tr}
\Big[{\frac{1}{1- \theta \hat F}}\delta \theta {\hat F} -
{\frac{1}{1- \theta \hat F}}\theta{\hat F}\delta \theta{\hat F}\cr
&& - \,\delta\theta {\hat F}\Big] + \, \, {\mathcal{O}}(\partial{\hat F}) 
+ \hbox{total derivatives} \cr
&=& 0 + {\mathcal{O}}(\partial{\hat F}) + {\hbox{total derivatives}}
\end{eqnarray}
In the last step we have used the following identities:
\begin{eqnarray}
\partial_l~ {\sqrt{\hbox{det}(1-\theta \hat F)}}
&=& -\frac{1}{2} {\sqrt
{\hbox{det}(1-\theta \hat F)}}\left({\frac{1}{1- \theta \hat F}}
\right)^{j}~_{m}~
\theta^{m i}\partial_l {\hat F}_{ij} \cr
{\hat D}_l~ {\sqrt{\hbox{det}(1-\theta \hat F)}}
&=& -\frac{1}{2} {\sqrt
{\hbox{det}(1-\theta \hat F)}}\left({\frac{1}{1- \theta \hat F}}
\right)^{j}~_{m}~
\theta^{m i}{\hat D}_l {\hat F}_{ij} \cr
&&\qquad + {\mathcal{O}}(\partial_l\hat F {\hat
D}_l\hat F)
\end{eqnarray}
and
\begin{center}
$\delta\theta^{kl}(\partial_l {\hat A}_k + \hat D_l \hat A_k) = \delta
\theta^{kl} \hat F_{lk}$
\end{center}
Now let us turn to the next lower form coupling. 
\begin{eqnarray}
&&\delta \left[{\sqrt{\hbox{det}(1-\theta \hat F)}}\,\,\hat
F\frac{1}{1-\theta\hat F}\right]\cr
&=& -\frac{1}{2}{\sqrt{\hbox{det}(1-\theta \hat F)}}\hbox{Tr}\left[
{\frac{1}{1- \theta
\hat F}} \delta \theta \hat F + \frac{1}{1- \theta \hat F} \theta
\delta \hat F \right] \hat F\frac{1}{1-\theta\hat F} \cr
&+& {\sqrt{\hbox{det}(1-\theta \hat F)}}\Big[\delta\hat F\,
\frac{1}{1-\theta\hat F} + \hat F\frac{1}{1-\theta\hat F}\delta\theta
\hat F\frac{1}{1-\theta\hat F}\cr
&+& \hat F\frac{1}{1-\theta\hat F}\theta
\delta\hat F\frac{1}{1-\theta\hat F}\Big]
\end{eqnarray}
From here it is easy to show, after substituting the expression for
$\delta\hat F$ and doing similar manipulations as for the top form
case, that $\delta \left[{\sqrt{\hbox{det}(1-\theta \hat F)}}~~\hat
F\frac{1}{1-\theta\hat F}\right]$ is also zero up to total derivatives
and terms which are beyond the DBI approximation. Similarly, one can
show that 
\begin{eqnarray}
&&\delta \left[{\sqrt{\hbox{det}(1-\theta \hat F)}}\,\,\hat
F\frac{1}{1-\theta\hat F}\wedge \hat F\frac{1}{1-\theta\hat
F}\wedge\dots\wedge \hat F\frac{1}{1-\theta\hat F} \right] \cr
&&= 0 + {\mathcal O}(\partial\hat F) + \hbox{total derivatives} 
\end{eqnarray}
Using these and keeping the B-field fixed under the variation, we can
show that the proposed noncommutative Chern-Simons terms are
equivalent to the commutative ones in the DBI approximation.

\section{Conclusions}
We have obtained the couplings of the noncommutative branes to constant
Ramond-Ramond fields for both BPS and non-BPS branes of type II
superstring theories. For the case of BPS branes we have used the fact
that a noncommutative brane can be obtained as a classical
configuration of infinitely many lower dimensional branes. We have
also proposed these couplings for a generic value of the description
parameter and then used them to show that these couplings are
equivalent to the ones in the commutative $\Phi = B$ description in the
DBI approximation. 

For non-BPS branes we have proposed the couplings guided by the
requirement of background independence in $\Phi = -B$ description. We
showed how to obtain the couplings of a bunch of non-BPS branes to
Ramond-Ramond forms by condensing the noncommutative tachyon over level N
noncommutative soliton. The couplings obtained this way exactly match
with the ones found in the literature. The couplings presented here
have been verified to be consistent with T-duality in \cite{tatar}.

Finally we would like to mention how these couplings can be
generalized to the case of non-constant Ramond-Ramond fields
\cite{okawao, smnvstwo, micheliutwo}. For this one must take the
RR-fields to be a functional of transverse coordinates in
the action of N instantons. Then to obtain the coupling on a Dp-brane
one can follow the same procedure as for the BPS branes in the case of
constant RR-fields. Expanding the RR-field in a
nonabelian taylor series in the momentum space, one would get an open
Wilson line \cite{liu, dastrivedi}. The symmetrized trace prescription
for the matrices in the action of instantons would lead to the
smearing of the operators found in the case of constant RR fields over
this open Wilson line. Comparing these couplings with their
commutative counterparts one gets a bunch of interesting identities
relating commutative and noncommutative variables including a closed
form for the Seiberg-Witten map. An application of these results in
finding an infinite subset of derivative corrections to both
commutative DBI and Chern-Simons actions in the Seiberg-Witten limit
can be found in \cite{sdsmnvs}.

\newpage

\bibliographystyle{amsalpha}

\begin{thebibliography}{A}

\bibitem{talklink} N.V. Suryanarayana, \textit{``Ramond-Ramond
Couplings of Noncommutative Branes''},
{\tt http://theory.tifr.res.in/strings/Proceedings/nemani/}.

\bibitem{noncommrefs} A. Connes, M. R. Douglas and A. Schwarz,
\textit{``Noncommutative Geometry and Matrix Theory: Compactification 
on Tori''}, hep-th/9711162, JHEP {\bf 9802}, 003 (1998)\\
M. Douglas and C. Hull, \textit{``D-branes and the 
Noncommutative Torus''}, hep-th/9711165, JHEP {\bf 9802}, 008 
(1998)\\
F. Ardalan, H. Arfaei and M. M. Sheikh-Jabbari,
\textit{``Noncommutative Geometry from Strings and Branes''},
hep-th/9810072, JHEP {\bf 9902}, 016 (1999)\\
C. Chu and P. Ho,
\textit{``Noncommutative Open String and D-brane''},
hep-th/9812219, Nucl.\ Phys.\ B {\bf 550}, 151 (1999)\\
V. Schomerus, \textit{``D-branes and Deformation
Quantization''}, hep-th/9903205, JHEP {\bf 9906}, 030 (1999).

\bibitem{SW} N. Seiberg and E. Witten, \textit{``String theory and
noncommutative geometry''}, JHEP{\bf 9909}, 032 (1999)
[hep-th/9908142].

\bibitem{cornalba} L. Cornalba and R. Schiappa, \textit{ ``Matrix
Theory Star Products from the Born-Infeld Action''}, hep-th/9907211,\\
L. Cornalba, \textit{ ``D-brane Physics and Noncommutative Yang-Mills
Theory''}, hep-th/9909081, \\
N. Ishibashi, \textit{``A Relation Between
Commutative and Noncommutative Descriptions of D-branes''},
hep-th/9909176.

\bibitem{seibnew} N. Seiberg, \textit{``A Note on Background
Independence in Noncommutative Gauge Theories, Matrix Model and
Tachyon Condensation''}, hep-th/0008013.

\bibitem{myers} M. Van Raamsdonk and W. Taylor, \textit{``Multiple
Dp-branes in Weak Background Fields''}, hep-th/9910052;
Nucl.Phys. {\bf B573} (2000) 703,\\
R. C. Myers, \textit{``Dielectric-Branes''}, hep-th/9910053,
JHEP {\bf 9912}, 022 (1999),\\
W. Taylor, \textit{ ``The M(atrix) Model of M
Theory''}, hep-th/0002016.

\bibitem{gmsone} R. Gopakumar, S. Minwalla and A. Strominger,
\textit{``Noncommutative solitons''}, hep-th/0003160; JHEP {\bf 05}
(2000) 020.

\bibitem{dmr} K. Dasgupta, S. Mukhi and G. Rajesh,
\textit{``Noncommutative Tachyons''}, hep-th/0005006; JHEP {\bf 06}
(2000) 022.

\bibitem{hklm} J. Harvey, P. Kraus, F. Larsen and E. Martinec,
\textit{``D-Branes and Strings as Noncommutative Solitons''},
hep-th/0005031, JHEP {\bf 07} (2000) 042.

\bibitem{hkl} J. Harvey, P. Kraus and F. Larsen, \textit{``Tensionless
Branes and Discrete Gauge Symmetry''}, hep-th/0008064.

\bibitem{sochichiu} C. Sochichiu, \textit{``Noncommutative Tachyonic
Solitons. Interaction with Gauge Field''}, hep-th/0007217; JHEP {\bf
08} (2000) 026.

\bibitem{gmstwo} R. Gopakumar, S. Minwalla and A. Strominger,
\textit{``Symmetry Restoration and Tachyon Condensation in Open String
Theory''}, hep-th/0007226.

\bibitem{sennew} A. Sen, \textit{``Some Issues in Noncommutative
Tachyon Condensation''}, hep-th/0009038.

\bibitem{sencs} A. Sen, \textit{``Supersymmetric World-Volume Action
For Non-BPS D-Branes''}, hep-th/9909062; JHEP {\bf 10} (1999) 008.

\bibitem{billo} M. Bill\`o, B. Craps and F. Roose,
\textit{``Ramond-Ramond Coupling of Non-BPS D-Branes''},
hep-th/9905157; JHEP {\bf 06} (1999) 033.

\bibitem{kennedy} C. Kennedy and Wilkins, \textit{``Ramond-Ramond
Couplings on Brane-Anti-Brane Systems''}, hep-th/9905195;
Phys. Lett. {\bf B464} (1999) 206.

\bibitem{twosens} B. Janssen and P. Meessen, \textit{``A Nonabelian
Chern-Simons Term for Non-BPS D-Branes''}, hep-th/0009025.

\bibitem{garousi} M.R. Garousi, \textit{``Tachyon Couplings on Non-BPS
D-Branes and Dirac-Born-Infeld Action''}, hep-th/0003122;
Nucl.Phys. {\bf B584} (2000) 284.

\bibitem{nccs} S. Mukhi and N.V. Suryanarayana, \textit{``Chern-Simons 
Terms on Noncommutative Branes''}, hep-th/0009101, 
JHEP {\bf 0011}, 006 (2000).

\bibitem{morariu} D. Brace, B. Morariu and B. Zumino,
\textit{``T-Duality and Ramond-Ramond Backgrounds in the 
Matrix Model''}, hep-th/9811213,
Nucl.Phys. {\bf B549} 181 (1999).

\bibitem{tatar} R. Tatar, \textit{``T-Duality and Actions for
Noncommutative D-branes''}, hep-th/0011057.

\bibitem{liu} H. Liu, \textit{``~$*$-Trek II: $*_n$ Operations, Open
Wilson Lines and the Seiberg-Witten Map''}, hep-th/0011125.

\bibitem{dastrivedi} S.R. Das and S. Trivedi, \textit{ ``Supergravity
Couplings to Noncommutative Branes, Open Wilson Lines and Generalized
Star Products''}, hep-th/0011131, JHEP {\bf 0102}, 046 (2001).

\bibitem{okawao} Y. Okawa and H. Ooguri, \textit{``An Exact Solution to 
Seiberg-Witten Equation of Noncommutative Gauge Theory''},
hep-th/0104036.

\bibitem{smnvstwo} S. Mukhi and N.V. Suryanarayana,
\textit{``Gauge-invariant Couplings of Noncommutative Branes to
Ramond-Ramond Backgrounds''}, hep-th/0104045, JHEP {\bf 0105}, 023
(2001).

\bibitem{micheliutwo} H. Liu and J. Michelson, \textit{``Ramond-Ramond 
Couplings of Noncommutative D-Branes''}, hep-th/0104139.

\bibitem{sdsmnvs} S.R. Das, S. Mukhi and N.V. Suryanarayana,
\textit{Derivative Corrections from Noncommutativity}, hep-th/0106024.

\end{thebibliography}

\end{document}